\documentclass[small extended]{svjour3}
\usepackage{amsmath}%
\usepackage{amsfonts}%
\usepackage{amssymb}%
\usepackage{graphicx}

\DeclareMathAlphabet{\mathpzc}{OT1}{pzc}{m}{it}

\smartqed 
\begin{document}
\title{Electromagnetic sources distributed on shells in a Schwarzschild background}

\author{N. {G{\"u}rlebeck} \and {J. {Bi{\v c}{\'a}k}}\and\\  A. C. Guti\'errez-Pi\~{n}eres}
\institute{
N. {G{\"u}rlebeck} \and {J. {Bi{\v c}{\'a}k}}
\at Institute of Theoretical Physics, Charles University, Prague, Czech Republic\\
\email{norman.guerlebeck@gmail.com}\\ 
\email{Jiri.Bicak@mff.cuni.cz}
\and
A. C. Guti\'errez-Pi\~{n}eres 
\at Facultad de Ciencias B\'asicas, Universidad Tecnol\'ogica de Bol\'ivar, Cartagena de Indias, Colombia\\
\email{acgutierrez@unitecnologica.edu.co}
}

\date{Received: date / Accepted: date}
\journalname{General Relativity and Gravitation}
\maketitle
\begin{abstract}
In the Introduction we briefly recall our previous results on stationary electromagnetic fields on black-hole backgrounds and the use of spin-weighted spherical harmonics. We then discuss static electric and magnetic test fields in a Schwarzschild background using some of these results. As sources we do not consider point charges or current loops like in previous works, rather, we analyze spherical shells with smooth electric or magnetic charge distributions as well as electric or magnetic dipole distributions depending on both angular coordinates. Particular attention is paid to the discontinuities of the field, of the 4-potential, and their relation to the source.
\end{abstract}

\section{Introduction}

J. S. Bach's ``Goldberg'' variations represent the beginning of the theme of musical variations followed by works of Beethoven, Brahms, Reger, and many others, most recently by a ``Bearbeitung'' of Bach's original by D. Sitkovetsky for the string trio and for the string orchestra. J. N. Goldberg's own ``Goldberg variations'' on the themes of equations of motion, conservation laws and gravitational radiation were among the first in the 1950-1960's which started the revival of general relativity.

Here we deal with a much simpler problem -- sources which are at rest. Still, we use the highly quoted work \cite{Goldberg_1967} by Goldberg and his colleagues in the Syracuse University on the spin-weighted spherical harmonics. We are happy to dedicate this note to Professor Goldberg's 86th birthday. However, we would also like to recall another anniversary: in April 2011 it will be 100 years after Albert Einstein came to Prague to spend 16 months at the German part of the Charles University. In 1912 Einstein was followed by P. Frank whose student who ``received much of his training with Philipp Frank in Prague before coming to the USA as Einstein's assistant'' \cite{Newman_2002} was Peter Bergmann. It was Frank who recommended him to Einstein. As is well-known, P.G. Bergmann founded the relativity group in Syracuse and as E.T. Newman writes in \cite{Newman_2002} J.N. Goldberg became Bergmann's first PhD student there. Is there not a clear connection between both anniversaries?

One of us used spin-weighted spherical harmonics extensively in several works. In \cite{Bicak_1975} we applied the Newman-Penrose (NP) formalism to develop an approximation procedure suitable for treating radiation problems, including wave tails, in non-linear electrodynamics. We also found conserved quantities, analogous to those discovered by Newman and Penrose in Maxwell's and Einstein's theories (cf. e.g. \cite{Newman_1965}) and analyzed, among others, by Goldberg \cite{Glass_1970,Goldberg_1972}. However, a deeper physical meaning of these quantities in, say, Born-Infeld non-linear electrodynamics remains to be seen.

The spin-weighted spherical harmonics and their generalization to spin-weighted spheroidal harmonics were crucial in the fundamental contribution by Teukolsky \cite{Teukolsky_1973} in which the equations for perturbations of the Kerr black holes were decoupled and separated. Some time ago we systematically considered stationary electromagnetic perturbations of the Schwarzschild black holes \cite{Bicak_1977} as well as of the Kerr black holes \cite{Bicak_1976}. We also analyzed in detail coupled electromagnetic and gravitational perturbations of the Reissner-Nordstr\"om black holes \cite{Bicak_1979,Bicak_1980}. In case of all these black holes we found general stationary vacuum solutions\footnote{In \cite{Bicak_1978} we also considered scalar perturbations of the Kerr-Newman black holes and determined stationary solutions.} and gave explicit solutions for fields of a number of special sources, like point charges and current loops in various positions outside the black holes. Stationary electromagnetic fields around black holes have later been used in various contexts in relativistic astrophysics, in particular in black-hole electrodynamics, \cite{Thorne_1986,Punsly_2008}, and in purely theoretical problems like discovering the Meissner effect for extremal objects in 3+1 and also in higher dimensions (see, e.g., \cite{Bicak_2000,Chamblin_1998}).

Recently, we were interested in the spherical gravitating condensers in general relativity \cite{Bicak_2010} -- two concentric shells made of perfect fluids restricted by the condition that the electric field is non-vanishing only between the shells. We used Israel's formalism and took energy conditions into account. When the shells approach each other while the total charge on a shell increases a sphere from spherical dipoles forms. However, in this process the energy condition cannot be satisfied, since the field between the shells becomes singular not even permitting to write down the energy-momentum tensor using classical distributions. A possibility to circumvent these problems is to consider test dipoles as we do in the present paper.

In this context we realized that we are not aware of an example of a surface distribution of dipoles in a curved background, or even of a general surface distribution of (monopole) test charges which do not share the symmetry of the background. In this note we construct simple examples of charges and dipoles distributed on spherical surfaces (shells) in a Schwarzschild background. We use the general vacuum stationary electromagnetic fields on the Schwarzschild background given in \cite{Bicak_1977} and formulate appropriate junction conditions. We are also interested in dipole distributions on the spheres, so we have to calculate 4-potentials of the vacuum fields, which was not done in the original work \cite{Bicak_1977}. Since here we discuss distributions on spherical surfaces in the spherically symmetric background, the results are neat and simple, resembling the results in classical electrodynamics in flat spacetime. In a future paper \cite{Gurlebeck_2011}, we turn to general dipole layers in general spacetimes.

\section{Stationary fields in a Schwarzschild background}\label{sec:Preliminaries}

We discuss solutions of electromagnetic test fields in a Schwarzschild background with curvature coordinates $(x^\mu)=(t,r,\theta,\varphi)$:
\begin{align}
\mathrm{d}s^2 = \left(1 - \frac{2M}{r}\right){\mathrm d}t^2 - \left(1 - \frac{2M}{r}\right)^{-1}{\mathrm d}r^2 - r^2\left({\mathrm d}\theta^2 + \sin^2\theta {\mathrm d}\varphi^2 \right).
\end{align}
We also allow magnetic sources in order to be able to describe magnetic dipole shells, so the Maxwell equations read\footnote{\label{footnote:convention}Note, that the signature of the metric and the orientation of the volume form are taken as in \cite{Jackson_1998}, with the important difference that our Maxwell tensor $F_{\alpha\beta}$ is the negative of the one defined there, i.e. the indices are exchanged.}
\begin{align}\label{eq:MaxwellEquation}
 F^{\alpha\beta}_{\phantom{\alpha\beta};\beta}=4\pi  j_{(e)}^\alpha,\quad {}^*F^{\alpha\beta}_{\phantom{\alpha\beta};\beta}=4\pi  j_{(m)}^\alpha,
\end{align} 
where the indices $(e/m)$ distinguish the electric/magnetic quantities. We also include magnetic monopoles for mathematical purposes since this will allow us to describe charge distributions depending on $\varphi$ in an easy way.  In the absence of either magnetic or electric sources an electric 4-potential $F_{\alpha\beta}=A^{(e)}_{\beta,\alpha}-A^{(e)}_{\alpha,\beta}$ or a magnetic 4-potential ${}^*F_{\alpha\beta}=A^{(m)}_{\beta,\alpha}-A^{(m)}_{\alpha,\beta}$ can be introduced. However, we will solve the Maxwell equations for the field directly and only afterwards integrate the fields to obtain the 4-potentials in order to verify jump conditions. It is useful to introduce the complex 4-current $J^\alpha$, 4-potential $\mathcal A_\alpha$, and Maxwell tensor $\mathcal F_{\alpha\beta}$
\begin{align}\label{eq:complexquantities}
J^\alpha=  j_{(e)}^\alpha+\mathrm{i}j_{(m)}^\alpha,\quad \mathcal A_\alpha=  A^{(e)}_\alpha+\mathrm{i} A^{(m)}_\alpha,\quad \mathcal F_{\alpha\beta}=F_{\alpha\beta}+\mathrm{i} {}^*F_{\alpha\beta}.
\end{align}
The equations for general electromagnetic test fields in a Kerr background were given within the NP formalism in the well-known paper by Teukolsky in \cite{Teukolsky_1973}. Using a separation technique, relying heavily on the spin-weighted spherical harmonics introduced by Goldberg \emph{et al.} in \cite{Goldberg_1967}, general stationary solutions were given as series expansions for sources on the Schwarzschild background in \cite{Bicak_1977}.
Before we discuss our solutions we repeat briefly some of the formalism we are going to need and refer the reader for details to \cite{Bicak_1977,Bicak_1976}. In the Schwarzschild spacetime the NP null tetrad used is given by
\begin{align}\label{eq:nulltetrad}
\begin{split}
 l^{\alpha}&=\left(\left(1-\frac{2M}{r}\right)^{-1},1,0,0\right),\quad n^{\alpha}=\frac 1 2 \left(1,-\left(1-\frac{2M}{r}\right),0,0\right),\\
 \quad m^{\alpha}&=\frac{1}{\sqrt{2}r}\left(0,0,1,\frac{\mathrm{i}}{\sin\theta}\right), \quad\, {\overline m}^{\alpha}=\frac{1}{\sqrt{2}r}\left(0,0,1,\frac{-\mathrm{i}}{\sin\theta}\right),
\end{split}
\end{align}
where a bar denotes complex conjugation. The Maxwell tensor $F_{\alpha\beta}$ expressed in terms of the three complex NP quantities $\phi_0,~\phi_1,~\phi_2$ reads\footnote{In contrast to \cite{Bicak_1977,Bicak_1976}, we use here $A_{[\alpha\beta]}=\tfrac{1}{2}(A_{\alpha\beta}-A_{\beta\alpha})$.}
\begin{align}\label{eq:definitionMaxwelltensor}
  F_{\alpha\beta}=4\mathrm{Re} \left[\phi_0 \overline m_{[\alpha}n_{\beta]}+\phi_1 (n_{[\alpha}l_{\beta]}+m_{[\alpha}\overline m_{\beta]}) +\phi_2 l_{[\alpha}m_{\beta]}\right].
\end{align}
Thus from the solutions for $\phi_i$, given explicitly below in Eqs. \eqref{eq:phis}, we can calculate the field $F_{\alpha\beta}$ and afterwards the four potential $A^{(e/m)}_\alpha$. Taking into account that the coefficients of the $\phi_i$ in \eqref{eq:definitionMaxwelltensor} are self-dual, i.e., they satisfy ${}^*A_{\alpha\beta}=-\mathrm{i} A_{\alpha\beta}$, we rewrite \eqref{eq:definitionMaxwelltensor} in the complex form
\begin{align}\label{eq:selfdualMaxwelltensor}
{\mathcal F}_{\alpha\beta}=4\left(\phi_0 \overline m_{[\alpha}n_{\beta]}+\phi_1 (n_{[\alpha}l_{\beta]}+m_{[\alpha}\overline m_{\beta]}) +\phi_2 l_{[\alpha}m_{\beta]}\right).
\end{align}
Hence, to obtain the entire information about the field it is sufficient to study just the time-space components of the self-dual tensor $\mathcal F_{\alpha\beta}$ given by
\begin{align}\label{eq:timecomponent}
\begin{split}
{\mathcal F}_{tr}=&-2\phi_1,\\
{\mathcal F}_{t\theta}=&\frac{1}{\sqrt 2}\left(\left(r-2 M\right) \phi_0-2 r \phi_2\right),\\
{\mathcal F}_{t\varphi}=&-\frac{\mathrm i}{\sqrt 2}\left(\left(r-2 M\right) \phi_0+2 r \phi_2\right)\sin\theta.
\end{split}
\end{align}

Let $r_1$ be the supremum and $r_2$ the infimum of the radii, such that points $(t,r,\theta,\varphi)$ lying in the support of $J^\alpha$ have $r\in(r_1,r_2)$. The electromagnetic field of a source with $2M < r_1 < r_2 < \infty$ is given in the region $2M \leq r < r_{_{1}}$ by
\begin{subequations}\label{eq:phis}
\begin{align}
\begin{split}
\phi_{0}&=\phantom{-}\sum\limits_{l,m}a_{lm}\sqrt{\frac{8}{l(l+1)}}\,{}_2F_1\left(1-l,l+2;2;\frac{r}{2M}\right) \;{}_{1}Y_{lm} ( \theta,\varphi ),\\
\phi_{1}&=\phantom{-}\sum\limits_{l,m}a_{lm}\,{}_2F_1\left(1-l,l+2;3;\frac{r}{2M}\right)\;{}_{0}Y_{lm}( \theta,\varphi ),\\
\phi_{2}&=-\sum\limits_{l,m}a_{lm}\sqrt{\frac{8}{l(l+1)}}\frac{M}{r}\, {}_2F_1\left(-l,l+1;2;\frac{r}{2M}\right)\;{}_{-1}Y_{lm} ( \theta,\varphi ),
\end{split}
\end{align}
whereas for $r > r_{_{2}}$ it reads
\begin{align}
\begin{split}
\phi_{0}&=-\sum\limits_{l,m}b_{lm}\sqrt{\frac{2l}{l+1}}\left(-\frac{2M}{r}\right)^{l+2}{}_2F_1\left(l+1,l+2;2l+2;\frac{2M}{r}\right)\;{}_{1}Y_{lm}(\theta,\varphi),\\
\phi_{1}&=\sum\limits_{l,m}b_{lm}\left(-\frac{2M}{r}\right)^{l+2}{}_2F_1\left(l,l+2;2l+2;\frac{2M}{r}\right) \;{}_{0}Y_{lm}(\theta,\varphi)+ \frac{Q}{2 r^2},\\
\phi_{2}&=-\sum\limits_{l,m}b_{lm}\sqrt{\frac{l}{2(l+1)}}\left(-\frac{2M}{r}\right)^{l+2}{}_2F_1\left(l+1,l;2l+2;\frac{2M}{r}\right)\;{}_{-1}Y_{lm}(\theta,\varphi),
\end{split}
\end{align}
\end{subequations}
where ${}_2F_1$ denotes a $(2,1)$-hypergeometric function, ${}_{s}Y_{lm}$ is a spin-s-weighted spherical harmonic, cf. \cite{Goldberg_1967}, and $\sum\limits_{l,m}$ is an abbreviation for $\sum\limits_{l=1}^{\infty}\sum\limits_{m=-l}^{l}$ (see \cite{Bicak_1977} for the detailed calculations). The constant $Q=Q^{(e)}+{\mathrm i} Q^{(m)}$ is the complex combination of the total electric charge $Q^{(e)}$ and magnetic charge $Q^{(m)}$. The constants $a_{lm}$ and $b_{lm}$ are also determined by the source via
\begin{align}\label{eq:constants}
\begin{split}
a_{lm} =&  \int\limits_{2M}^{\infty}\int\limits_{0}^{2\pi}\int\limits_{0}^{\pi}J_{lm}(r,\theta,\varphi )\,{}_2F_1\left(l,l+2;2l+2;\frac{2M}{r}\right)\left(-\frac{2M}{r}\right)^{l-2} {\mathrm d}\theta{\mathrm d}\varphi{\mathrm d}r,\\
b_{lm} =& \int\limits_{2M}^{\infty}\int\limits_{0}^{2\pi}\int\limits_{0}^{\pi}J_{lm}(r,\theta,\varphi)\,{}_2F_1\left(1-l,l+2;3;\frac{r}{2M}\right)\left(\frac{r}{2M}\right)^4{\mathrm d}\theta{\mathrm d}\varphi{\mathrm d}r,
\end{split}
\end{align}
where $J_{lm}$ -- vanishing for $r\notin(r_1,r_2)$ -- is defined in terms of the complex 4-current $J_\alpha$ as follows, cf. \eqref{eq:complexquantities}:
\begin{align}\label{eq:transformedcurrent}
\begin{split}
 J_{lm}(r,\theta,\varphi )=-\kappa_{lm}&\sin\theta \overline{Y}_{lm}(\theta,\varphi )\left[\left(\sqrt{8}r\cot\frac{\theta} 2{\overline m}^{\alpha}-\frac 2 r n^{\alpha}\right)J_{\alpha}\right.\\
  +&m^{\alpha}\left({\overline m}^{\beta}J_{\beta}\right)_{,\,\alpha}-\left.l^{\alpha}\left({n}^{\beta}J_{\beta}\right)_{,\,\alpha}\right].
\end{split}
\end{align}
The constants $\kappa_{lm}$ are given by $\tfrac{4\pi M \left[(l + 1)!\right]^2}{(2l + 1)!}$.

In vacuum regions we can introduce a complex 4-potential $\mathcal A_\alpha$. For stationary sources we can integrate $\mathcal F_{\alpha\beta}=\mathcal A_{\beta,\alpha}-\mathcal A_{\beta,\alpha}$ for the time component $\mathcal A_t$.
Using equations \eqref{eq:timecomponent} and \eqref{eq:phis} we obtain
\begin{align}\label{eq:potential}
r<r_1 :\,\, \mathcal A_t=&-\sum\limits_{l,m}a_{lm}\frac{8M}{l(l+1)}\,{}_2F_1\left(-l,l+1;2;\frac{r}{2M}\right)Y_{lm}(\theta,\varphi)+\mathcal A_{t0-},\notag\\
r>r_2 :\,\, \mathcal A_t=&\sum\limits_{l,m} b_{lm}\frac{4M}{l+1}\left(-\frac{2M}{r}\right)^{l+1}{}_2F_1\left(l,l+1;2l+2;\frac{2M}{r}\right)Y_{lm}(\theta,\varphi)\notag\\
&+\frac{Q}{r}+\mathcal A_{t0+}.
\end{align}
The constants $\mathcal A_{t0\pm}$ represent the remaining gauge freedom of the potential. 
These potentials are new - they were not calculated in \cite{Bicak_1977,Bicak_1976} since the field contains usually all needed information, however, for discussing dipole shells, it is important to know the potentials. For completeness we give also the remaining components of the 4-potential obtained by a separation ansatz. They read for $r<r_1$
\begin{align}
\mathcal A_r=&-\sum\limits_{l,m}a_{lm} \frac{4M r}{r-2M} \,{}_2F_1\left(-l,l+1;2;\frac{r}{2M}\right)g_{lm}(\theta)\mathrm{e}^{\mathrm{i}m\varphi},\notag\\
\mathcal A_\theta=&\sum\limits_{l,m}R^{(i)}_{lm}(r)\left(g_{lm,\theta}(\theta)\mathrm{e}^{\mathrm{i}m\varphi}+\frac{{}_{1}Y_{lm}(\theta,\varphi)+{}_{-1}Y_{lm}(\theta,\varphi)}{\sqrt{l(l+1)}}\right),\notag\\
\mathcal A_\varphi=&-\mathrm{i}\sum\limits_{l,m}R^{(i)}_{lm}(r)\left(-mg_{lm}(\theta)\mathrm{e}^{\mathrm{i}m\varphi}+\frac{{}_{1}Y_{lm}(\theta,\varphi)-{}_{-1}Y_{lm}(\theta,\varphi)}{\sqrt{l(l+1)}}\right),\notag\\
R^{(i)}_{lm}(r)=&a_{lm} r^2 \,{}_2F_1\left(1-l,l+2;3;\frac{r}{2M}\right),
\end{align}
and for $r>r_2$
\begin{align}
\mathcal A_r=&-\sum\limits_{l,m}b_{lm} \frac{4M^2 l}{r-2M} \left(-\frac{2M}{r}\right)^{l}\,{}_2F_1\left(-l,l+1;2;\frac{r}{2M}\right)h_{lm}(\theta)\mathrm{e}^{\mathrm{i}m\varphi},\notag\\
\mathcal A_\theta=&\sum\limits_{l,m}R^{(e)}_{lm}(r)\left(h_{lm,\theta}(\theta)\mathrm{e}^{\mathrm{i}m\varphi}+\frac{{}_{1}Y_{lm}(\theta,\varphi)+{}_{-1}Y_{lm}(\theta,\varphi)}{\sqrt{l(l+1)}}\right),\notag\\
\mathcal A_\varphi=&-\mathrm{i}\sum\limits_{l,m}R^{(e)}_{lm}(r)\left(-mh_{lm}(\theta)\mathrm{e}^{\mathrm{i}m\varphi}+\frac{{}_{1}Y_{lm}(\theta,\varphi)-{}_{-1}Y_{lm}(\theta,\varphi)}{\sqrt{l(l+1)}}\right)+\notag\\
&\mathrm{i}Q\cos\theta,\notag\\
R^{(e)}_{lm}(r)=&b_{lm} 4M^2 \left(-\frac{2M}{r}\right)^l \,{}_2F_1\left(l,l+2;2 l+2;\frac{2M}{r}\right).
\end{align}
The arbitrary functions $g_{lm}(\theta)$ and $h_{lm}(\theta)$ give some of the gauge freedom, e.g., the Lorenz gauge is achieved in the vacuum region with the choice $g_{lm}(\theta)=h_{lm}(\theta)=0$. In general, we could have assumed a $t$-dependent and a general $\varphi$-dependent gauge function.

\section{A direct approach for spherical shells}\label{sec:directapproach}

We now apply the solutions described above to the sources characterized by a 4-current
\begin{align}\label{eq:source}
{J}_{lm}^{\alpha}=f_{lm}(r)Y_{lm}(\theta,\varphi)\xi^\alpha,
\end{align}
where indices $l$ and $m$ label the source, $Y_{lm}$ denotes a spherical harmonic and $\xi^{\alpha}$ is the timelike Killing vector of the Schwarzschild metric. If $m=0$ the 4-current is real, i.e., it consists just of an electric part. If it is complex one has also magnetic parts. In order to obtain a 4-current which is proportional to $\xi^\alpha$ like in \eqref{eq:source} we need a stationary source, i.e., at least the spatial components of the net current must vanish. Nevertheless, two components which, for example, counter-rotate would also be admissible but we restrict ourselves here to a source which is at rest with respect to the Schwarzschild coordinates. Such sources consist of electric and magnetic parts. For such sources the sums, like in \eqref{eq:phis}, reduce to a single term and the calculations can be handled more easily. These sources form a complete set over the unit sphere which allows us to generalize the results to arbitrary sources. 

If a purely electric (magnetic) stationary source is considered, only an electric (magnetic) field arises which amounts to taking the real (imaginary) part of the right-hand sides of equation \eqref{eq:timecomponent}. This means taking a combination of fields produced by currents $J^\alpha$ and $\overline{J}^\alpha$.

Although most of our results hold for a general $f_{lm}(r)$, we concentrate in particular on spherical thin shells with radius $r_0$ covered by generally distributed electric/magnetic charge densities or by electric/magnetic dipole densities. The sources of such shells are respectively given by
\begin{align}
\begin{split}
  J_{\mathpzc{Mo}}^{\alpha}&=s_{\mathpzc{Mo}}^a\mathrm{e}_a^\alpha \left(1 - \frac{2M}{r_0}\right)^{\frac1 2}\delta(r-r_0),\\
  J_{\mathpzc{Di}}^{\alpha}&=s_{\mathpzc{Di}}^a \mathrm{e}_a^\alpha   \left(1 - \frac{2M}{r_0}\right) \frac{r_0^2}{r^2} \delta'(r-r_0).
\end{split}
\end{align}
We give a detailed derivation\footnote{Note that the definition of $\delta$-distribution is such that for any test function $f$ the integral over a spacetime region $\Omega$ in the Schwarzschild coordinates yields
$\int\limits_\Omega f(t,r,\theta,\varphi) \left(1 - \frac{2M}{r_0}\right)^{\frac1 2}\delta(r-r_0)\mathrm{d}\Omega=\int\limits_{\Sigma\cap \Omega}f(r_0,t,\theta,\varphi)\mathrm{d}\Sigma$, where $\mathrm d \Omega=\sqrt{-g}\mathrm{d}t\mathrm{d}r\mathrm{
d} \theta\mathrm{d}\varphi$ is the volume element of the spacetime and $\mathrm{d}\Sigma$ is the volume element of the hypersurface $\Sigma$ respectively. For the normal derivative of the $\delta$-distribution we obtain $\int\limits_\Omega f(t,r,\theta,\varphi)  
\left (1 - \frac{2M}{r_0}\right) \frac{r_0^2}{r^2} \delta'(r-r_0)\mathrm{d}\Omega=-\int\limits_{\Sigma\cap \Omega}(n^\mu f_{,\mu})(r_0,t,\theta,\varphi)\mathrm{d}\Sigma$ with the unit normal $n^\mu$ pointing outwards.}
of the exact form of $J_{\mathpzc{Di}}^{\alpha}$ including general backgrounds and non-stationary sources in another paper \cite{Gurlebeck_2011}. 
The vectors $\mathrm{e}_a^{\alpha}=\tfrac{\partial x^\alpha}{\partial\zeta^a}$ denote the tangential vectors to the hypersurface $\Sigma$ which represents the history of the shell. Using $(\zeta^a)=(t,\varphi,\theta)$ as intrinsic coordinates in $\Sigma$ these 
vectors become the coordinate vectors associated with the Schwarzschild coordinates; in particular, $\mathrm{e}_t^\alpha$ coincides with the Killing vector $\xi^\alpha$. The monopole surface current $s_{\mathpzc{Mo}}^a=s_{\mathpzc{Mo}}^{(e)a}+{
\mathrm i}s_{\mathpzc{Mo}}^{(m)a}$ as well as the dipole surface current $s_{\mathpzc{Di}}^a=s_{\mathpzc{Di}}^{(e)a}+{\mathrm i}s_{\mathpzc{Di}}^{(m)a}$ consist of an electric and a magnetic part. The surface currents can be written as
\begin{align}
s^a_{\mathpzc{Mo}}=\sigma u^a, \quad s^a_{\mathpzc{Di}}=d u^a,\quad \sigma=\sigma^{(e)}+\mathrm{i}\sigma^{(m)},\quad d=d^{(e)}+\mathrm{i}d^{(m)},
\end{align}
where $u^a$ is the velocity of the sources within the hypersurface $\Sigma$, $\sigma^{(e/m)}$ is the electric/magnetic rest surface charge density and $d^{(e/m)}$ the rest surface dipole density. For stationary sources we have  $u^a\mathrm{e}_a^\alpha\propto\xi^\alpha$ and thus $u^a=((1-\tfrac{2M}{r_0})^{-\tfrac 1 2},0,0)$. Hence, functions $f_{lm}$ in \eqref{eq:source} are given in the case of charges distributed on the shell by
\begin{align}\label{eq:generaltochargedcase}
 f^{\mathpzc{Mo}}_{lm}(r)=\hat\sigma_{lm} \delta(r-r_0),
\end{align}
where $\hat\sigma_{lm}$ are complex constants. Then the electric and magnetic charge densities become $\sigma_{lm}(\theta,\varphi)=\hat\sigma_{lm} Y_{lm}(\sigma,\varphi)$. Analogously, for shells covered by dipoles we obtain
\begin{align}\label{eq:generaltodipole}
\begin{split}
f^{\mathpzc{Di}}_{lm}(r)=\hat d_{lm}  \left(1 - \frac{2M}{r_0}\right)^{\frac 1 2} \frac{r_0^2}{r^2} \delta'(r-r_0),\quad d_{lm}(\theta,\varphi)=\hat d_{lm}Y_{lm}(\theta,\varphi).
\end{split}
\end{align}

Applying formulas \eqref{eq:nulltetrad}--\eqref{eq:transformedcurrent} given in Section \ref{sec:Preliminaries} to the sources \eqref{eq:source} we get for the Maxwell tensor for $l=m=0$ \footnote{The indices $l,m$ of the field $F_{\alpha\beta}$ and of the potential $A_{\alpha}$ are suppressed.}
\begin{align}\label{eq:field_l=0}
\begin{split}
r<r_1:\quad {\mathcal F}_{t\theta}&={\mathcal F}_{t\varphi}=0,\quad {\mathcal F}_{tr}=0,\\
r>r_2:\quad {\mathcal F}_{t\theta}&={\mathcal F}_{t\varphi}=0,\quad {\mathcal F}_{tr}=-\frac{Q}{r^2},
\end{split}
\end{align}
independently of $f_{00}(r)$. Of course, outside of a spherical symmetric distribution only the total charge is important, not its radial distribution. Note that this field coincides with the field obtained in electrostatics in flat space. Furthermore, if a spherical shell endowed with a constant dipole density is considered, i.e., $Q=0$ and \eqref{eq:source} holds together with \eqref{eq:generaltodipole}, there is no field present like in flat space. Therefore, the existence of such a dipole layer can be proven only by examining the trajectories of particles crossing it but not by measuring a distant field.

Given sources \eqref{eq:source} with a fixed $(l',m')$ with $l'>0$ and an arbitrary $f_{l'm'}$, the coefficients $a_{lm}$ and $b_{lm}$ vanish except for $(l,m)=(l',m')$. Assuming the radial distribution falls off sufficiently fast the coefficients read
\begin{align}\label{constants}
\begin{split}
a_{l'm'}&=-\frac{\kappa_{lm}}{8M^2}\int\limits_{2M}^{\infty}\tilde f_{l'm'}\frac{\mathrm d}{{\mathrm d}r}\left(\left(-\frac{2M}{r}\right)^{l}{}_2F_1\left(l,l+2;2l+2;\frac{2M}{r}\right)\right){\mathrm d}r,\\
b_{l'm'}&=-\frac{\kappa_{lm}}{8M^2}\int\limits_{2M}^{\infty}\tilde f_{l'm'} \frac{\mathrm d}{{\mathrm d}r}\left(\left(\frac{r}{2M}\right)^2 {}_2F_1\left(1-l,l+2;3;\frac{r}{2M}\right)\right){\mathrm d}r,\\
\tilde f_{l'm'}&=(r^2- 2Mr) f_{l'm'}(r).
\end{split}
\end{align}

For example, the field of the source with $(l,m)=(1,0)$ can be simplified for $r<r_1$ to read
\begin{align}
\begin{split}
F_{tr}=(E_0+\mathrm i B_0) \cos\theta,\quad F_{t\theta} =-(E_0+\mathrm i B_0) (r-2M)\sin\theta, 
\end{split}
\end{align}
where $E_0=-a_{10}\sqrt{\frac{3}{\pi}}$. This solution coincides with the standard asymptotically homogeneous electric and magnetic field. In the limit $r_1\to\infty$,  $J_{10}^{\alpha}$ provides a source of this field. In \cite{Gurlebeck_2011} we discuss a disc source generated by such a field.

\section{Discontinuities in the electric field and the potential}

What are the jumps caused in the field and the potential by sources of the form \eqref{eq:generaltochargedcase} and \eqref{eq:generaltodipole}? Whereas the results regarding the first are known in general in the electric case, see \cite{Kuchar_1968}, the jumps of the field of dipole layers were not analyzed before. After we obtain the jumps for these special sources we can superpose them to generalize the results to arbitrary charge and dipole distributions. 
The jump of a function $g$ across a spherical shell with radius $r_0$ is defined as $[g]=\lim\limits_{r\to r_0+}g(t,r,\theta,\varphi)-\lim\limits_{r\to r_0-}g(t,r,\theta,\varphi)$.

Since we look at stationary sources the electric charges do not produce magnetic fields, the jump in these are solely caused by magnetic charges. 
Therefore, we can assume that both kind of charges are present at the same time.
For such spherical shells we obtain from \eqref{eq:generaltochargedcase}, \eqref{eq:constants}, using some standard identities for hypergeometric functions, see e.g. \cite{Erdelyi_1953}, and Abel's identity, the following conditions:
\begin{align}\label{eq:jumpschargedensity}
[\mathcal F_{t\theta}]=[\mathcal F_{t\varphi}]=0,\quad [F_{tr}]=-4\pi\hat\sigma_{lm} Y_{lm}(\theta,\varphi).
\end{align} 
This resembles classical results -- the normal component of the electric (or magnetic) field jumps across a layer of electric (or magnetic) charges, whereas the tangential components are continuous. The jump is given by the corresponding charge density.
These jump conditions are in the form\footnote{The difference in sign is due to a different conventions explained in footnote \ref{footnote:convention}.} found in \cite{Kuchar_1968} if the coordinates $(\zeta^a)=(t,\theta,\varphi)$ are used as intrinsic coordinates and $n^{\mu}=\left(1-\tfrac{2M}{r_0}\right)^{\tfrac 1 2}\delta^\mu_r$ as the unit normal pointing outwards:
\begin{align}
\begin{split}
 [\mathcal F_{t\bot}]&=[\mathcal F_{\mu\nu}\mathrm{e}^\mu_t n^\nu]=-4\pi\hat\sigma_{lm} Y_{lm}(\theta,\varphi)\left(1-\frac{2M}{r_0}\right)^{\frac 1 2}=-4\pi s^{\mathpzc{Mo}}_t,\\
 [\mathcal F_{\theta\bot}]&=[\mathcal F_{\varphi\bot}]=0.
\end{split}
\end{align} 
These results can now be superposed to obtain an arbitrary charge density $\sigma(\theta,\varphi)=\sum\limits_{l=0}^\infty\sum\limits_{m=-l}^l \hat\sigma_{lm}Y_{lm}(\theta,\varphi)$ at the shell which results in the general jump conditions
\begin{align}\label{eq:jumpschargedensitygeneral}
[\mathcal F_{t\theta}]=[\mathcal F_{t\varphi}]=0,\quad [F_{tr}]=-4\pi\sigma(\theta,\varphi).
\end{align}
In order to discuss the jumps in the 4-potential it is sufficient to consider only the scalar potential $\mathcal A_t=\mathcal A_\mu\xi^\mu$ and, of course, this is continuous across the shell.

In the more interesting case of spherical shells endowed with a dipole density $\hat p_{lm} Y_{lm}(\theta,\varphi)$, analogous calculations lead to the relations
\begin{align}\label{eq:jumpsdipolelayer}
\begin{split}
 [\mathcal A_t]&=4\pi \hat d_{lm}\left(1-\frac{2M}{r_0}\right)^{\frac 1 2} Y_{lm}(\theta,\varphi),\\
 [\mathcal F_{t\theta}]&=-4\pi \left(1-\frac{2M}{r_0}\right)^{\frac 1 2}\hat d_{lm} \frac{\partial}{\partial\theta}Y_{lm}(\theta,\varphi),\\
 [\mathcal F_{t\varphi}]&=-4\pi \left(1-\frac{2M}{r_0}\right)^{\frac 1 2} \hat d_{lm} \frac{\partial}{\partial\varphi}Y_{lm}(\theta,\varphi),\\
 [\mathcal F_{tr}]&=0
\end{split}
\end{align}
where we used the gauge $A_{t0+}-A_{t0-}=4\pi d_{r0}Y_{00}\delta^0_l$, cf. \eqref{eq:potential} and the discussion after \eqref{eq:field_l=0}. They are again analogous to the conditions for a dipole layer in flat space.
These relations can also be generalized for an arbitrary dipole density on the sphere using the completeness of  spherical harmonics:
\begin{align}\label{eq:jumpsdipolelayergeneral}
\begin{split}
 [\mathcal A_t]&=4\pi d(\theta,\varphi)\left(1-\frac{2M}{r_0}\right)^{\frac 1 2}=4\pi s_t^{\mathpzc{Di}},\\
 [\mathcal F_{t\theta}]&=-4\pi \left(1-\frac{2M}{r_0}\right)^{\frac 1 2} \frac{\partial}{\partial\theta}d(\theta,\varphi)=-4\pi s^{\mathpzc{Di}}_{t,\theta},\\
 [\mathcal F_{t\varphi}]&=-4\pi \left(1-\frac{2M}{r_0}\right)^{\frac 1 2} \frac{\partial}{\partial\varphi}d(\theta,\varphi)=-4\pi s^{\mathpzc{Di}}_{t,\varphi},\\
[\mathcal F_{tr}]&=0.
\end{split}
\end{align}
The jumps of the tangential components are trivially obtained from the jump of the potential. Again, we have a situation like in the classical theory that the normal component of the field does not jump and the tangential components do jump, where the amount is given by the derivative of the surface current in the respective tangential direction, cf. \eqref{eq:jumpsdipolelayergeneral}. The jump of the potential is directly given by the surface current. 

In this paper the behavior of the discontinuities for spherical, static dipole shells was directly proven in the Schwarzschild spacetime. This was possible since the general solution to the Maxwell equation in this background is known.
Since in this note we wished to analyze both electrostatics and magnetostatics in a unified framework using complex quantities \eqref{eq:complexquantities}, we did not, for example, treat the case of electric currents moving along the shells and producing magnetic fields so that jump conditions like
\begin{align}
 [A^{(e)}_a]=4\pi s^{\mathpzc{Di}}_a
\end{align}
arise for moving electric dipoles. In a later paper \cite{Gurlebeck_2011}, general sources, arbitrary hypersurfaces and arbitrary spacetimes will be considered. 

In principle, it would also be possible to assume different mass parameters in the different portions of space, thereby generating a massive shell; in particular, introducing flat space inside the shell is of interest. The same analysis can be done in such a case but the time coordinate will not go continuously through the shell anymore. Other intrinsic coordinates, e.g. the proper time of an observer at rest in the shell, are necessary in such a case (cf., e.g., \cite{Bicak_2010}).

Knowing the junction conditions we can now use them in the following ``indirect'' approach: start out from some known metrics and fields, assume they are glued together and from the jumps deduce whether this procedure yields physically plausible sources on the junction. This procedure will be employed in our future work \cite{Gurlebeck_2011}.

\begin{acknowledgement}
We thank Tom\'a\v s Ledvinka for discussions. JB acknowledges the partial support from Grant No. GA\v CR 202/09/00772 of the Czech Republic, of Grants No. LC06014
and No. MSM0021620860 of the Ministry of Education. NG was financially supported by the PhD-student Grant  No. GAUK. 22708 and No. GA\v CR 205/09/H033. JB and NG are also grateful to the Albert Einstein Institute in Golm for the kind hospitality.
A. C. G-P. acknowledges the hospitality of the Institute of Theoretical Physics, Charles University (Prague) and the financial support from COLCIENCIAS, Colombia. 
\end{acknowledgement}

\end{document}